# Hybrid Optical Modes in Hexagonal Crystals

A. Dyson[a] and B. K. Ridley[b]

a School of Mathematics, Statistics & Physics, Newcastle University, UK.

b School of Computer Science and Electronic Engineering, University of Essex, Colchester, UK.

In nanostructure electronic devices, it is well-known that the optical lattice waves in the constituent semiconductor crystals have to obey both mechanical and electrical boundary conditions at an interface. The theory of hybrid optical modes, established for cubic crystals, is here applied to hexagonal crystals. In general, the hybrid is a linear combination of a longitudinally-polarized (LO) mode, an interface mode (IF), and an interface TO mode. It is noted that the dielectric and elastic anisotropy of these crystals add significant complications to the assessment of the electro-phonon interaction. We point out that, where extreme accuracy is not needed, a cubic approximation is available. The crucial role of lattice dispersion is emphasized. In the extreme long-wavelength limit, where lattice dispersion is unimportant, the polar optical hybrid consists of an LO component plus an IF component only. In his case no fields are induced in the barrier, and there are no remote-phonon effects.

1. **Introduction**

The electron-phonon interaction involving polar optical modes in a cubic crystal is well-known [1 – 4]. In the case of hexagonal crystals, description of the effect has been limited to the account given by the dielectric continuum (DC) modes [5 – 8], which ignore mechanical boundary conditions. In cubic crystals, taking mechanical boundary conditions into account (which is always the case for acoustic modes) requires the hybridization of longitudinal and transverse modes in a number of circumstances. In cubic crystals, the longitudinally polarized optical (LO) mode requires a linear combination with an electromagnetic interface (IF) mode, plus, if lattice dispersion is taken into account, an interface transversely-polarized optical (TO) mode, all components of the hybrid having the same frequency. Such a hybrid allows both mechanical and the usual electric boundary conditions to be satisfied. The aim is, in what follows, to apply hybridization theory to hexagonal crystals.

An understanding of the physics of polar-crystals crystallized over the last years of the twentieth century in books by Seitz (1940) [9], Mott and Gurney (1940) [10], Born and Huang (1954) [11], Haynes and Loudon (1978) [12] among others, and by the study of the Raman effect by Loudon (1964) [13]. The study of travelling electromagnetic waves in an orthorhombic polar crystal free of space charge and electric currents lead to a fundamental solution of Maxwell's equations:

$$-c^2 \boldsymbol{q}\boldsymbol{q}.\boldsymbol{E} + (c^2 q^2 - \omega^2 \boldsymbol{\kappa})\boldsymbol{E} = \boldsymbol{0} \qquad 1$$

c is the velocity of light in vacuo, **q** the wave vector of the wave of the form $\boldsymbol{E}e^{i(\boldsymbol{q}.\boldsymbol{R}-\omega t)}$, where E is the electric field, $\omega$ is the angular frequency and $\kappa$ is the dielectric constant, now a vector quantity. This is an equation for the components of the electric field, $E_x$, $E_y$, $E_z$. There are two sorts of solutions. One is for transversely-polarized waves (so-called ordinary waves):

$$(c^2 q^2 - \omega^2 \boldsymbol{\kappa})\boldsymbol{E} = \boldsymbol{0} \qquad 2$$

These describe waves of light, their speed determined by the components of the dielectric vector $\kappa_x, \kappa_y, \kappa_z$. The other solution (for so-called extraordinary waves, longitudinally polarized) is a set of three simultaneous equation for the field components. Solutions exist provided the relevant determinant vanishes i.e

$$c^4 q^2 (\kappa_x q_x^2 + \kappa_y q_y^2 + \kappa_z q_z^2)$$
$$-c^2 \omega^2 [\kappa_x(\kappa_y + \kappa_z) + \kappa_y(\kappa_z + \kappa_x) + \kappa_z(\kappa_x + \kappa_y)] + \kappa_x \kappa_y \kappa_z \omega^4 = 0 \qquad 3$$

Modes of interest have frequencies far below the light line, so the first term in eq, 3 dominates:

$$\kappa_x q_x^2 + \kappa_y q_y^2 + \kappa_z q_z^2 = 0 \qquad 4$$

Lattice dynamics provides the equations of motion for ionic displacement for polar modes:

$$\omega_{Li}^2 \mu u_i = \omega_{TOi}^2 \mu u_i - e_i^* E_i + \Omega_i(q^2) \qquad 5$$

where $\mu$ is the reduced ionic mass, $\omega_{Ti}$ is the frequency of the non-polar TO modes, $e^*$ is the effective charge associated with the macroscopic field, and $\Omega_i$ represents the elastic forces that describe the lattice dispersion ($i = x, y, z$). In what follows lattice dispersion will be ignored, corresponding to a long-wavelength approximation. Straightforward manipulation of the equation of motion and the electric field yields the following relations:

$$D_i = \epsilon_i(\omega) E_i = \epsilon_{\infty i} E_i + \epsilon_i^* u_i / V_0 \qquad 6$$

$$E_i = \frac{\epsilon_i^*}{V_0(\epsilon_i(\omega) - \epsilon_{\infty i})} u_i \qquad 7$$

$$\epsilon_i^{*2} = \omega_{Ti}^2 \mu V_0 (\epsilon_i(0) - \epsilon_{\infty i}) \qquad 8$$

$$\epsilon_i(\omega) = \epsilon_{\infty i} \frac{\omega^2 - \omega_{Li}^2}{\omega^2 - \omega_{Ti}^2} \qquad 9$$

Here, $\epsilon_i^*$ is the effective ionic charge, $\epsilon_i(\omega)$ is the frequency-dependent permittivity, $\epsilon_{\infty i}$ is the high-frequency permittivity, $\epsilon_{si}$ is the static permittivity, $u_i$ is the ionic displacement and $V_0$ is the volume of the unit cell. Following Loudon it is reasonable to take the high frequency permittivity to be isotropic, so $\epsilon_{\infty i} = \epsilon_\infty$, a scalar.

## 2. Dispersion in a hexagonal crystal

In a cubic crystal, dispersion, the relationship between frequency and wave vector of travelling waves is determined by the lattice elastic forces that describe the variation of frequency throughout the Brillouin zone. In a hexagonal crystal extra dispersion is provided by its polar character. Eq.4 becomes

$$\kappa_x q_x^2 + \kappa_z q_z^2 = 0 \qquad 10$$

where x is chosen to be a coordinate in the basal plane, considered to be isotropic, and z the coordinates along the optic axis. We need to consider the role of lattice dispersion and its effect on the dielectric properties of travelling waves and interface modes. The dispersion of LO and TO modes is of the form:

$$\omega_{Li}^2 = \omega_{LOi}^2 - \Omega_{Li}(q^2) \qquad 11$$

$$\omega_{Ti}^2 = \omega_{TOi}^2 - \Omega_{Ti}(q^2)$$

$i = x, z$. Note that eq.9 shows that dispersion does not alter the fact that for LO modes the permittivity is zero. The equation of motion for the LO modes is:

$$\omega_{Li}^2 \mu u_i = c_i u_i - e_i^* E_i - \Omega_{Li}(q^2) \qquad 12$$

$c_i$ is the ionic force constant related to the TO zone-centre frequency $c_i = \omega_{TOi}^2 \mu$. The zone-centre LO frequency is given from:

$$\omega_{LOi}^2 \mu = \omega_{TOi}^2 \mu - \frac{e_i^{*2}}{V_0(\epsilon(\omega_{Li})-\epsilon_\infty)} \rightarrow \omega_{Ti}^2 \mu + \frac{e_i^{*2}}{V_0 \epsilon_\infty} \qquad 13$$

The ionic charge (eq.8) can now be written:

$$e_i^{*2} = (\omega_{LOi}^2 - \omega_{TOi}^2)\mu V_0 \epsilon_{\infty i} \qquad 14$$

It is worth pointing out that, according to eq.9, the permittivity becomes a function of dispersion for arbitrary frequency. This dependency is a crucial factor in distinguishing between LO and IF permittivities. In the LO case, $q^2$ is real and positive; in the IF case q is imaginary and

$$q^2 = q_x^2 - q_z^2 \qquad 15$$

Because the velocity of light is very large, $q^2$ has to be very small for the IF frequency to be of order of the LO frequency. Thus, the permittivity associated with the IF mode is:

$$\epsilon_i(\omega) = \epsilon_{\infty i} \frac{\omega^2 - \omega_{LOi}^2}{\omega^2 - \omega_{TOi}^2} \qquad 16$$

A different sort of dispersion is implied by eq.10 When both $q_x$ and $q_z$ are real quantities, the only way this equation is satisfied is for $\kappa_x$ or $\kappa_z$ to be negative, which means that either $\omega^2 < \omega_{Lx}^2$ and $\omega^2 > \omega_{Tx}^2$, or $\omega^2 < \omega_{Lz}^2$ and $\omega^2 > \omega_{Tz}^2$. In the case of GaN, for example, $\kappa_z > \kappa_x$, so for $q^2 = q_x^2 + q_z^2$ the equation describes a band of frequencies $\omega_{Lx}^2 \leq \omega^2 \leq \omega_{Lz}^2$. Taking $\theta$ to be the angle between the direction of propagation and the z axis, we have:

$$\kappa_x \sin^2\theta + \kappa_z \cos^2\theta = 0 \qquad 17$$

Inserting the frequency dependence of the dielectric constant, we obtain the dispersion equation that relates $\omega^2$ and $q^2$. This is a quadratic in $\omega^2$ with two solutions. Using the useful fact that the difference between the two LO frequencies is very small compared to either frequency, and the same is true for the TO frequencies, the two solutions, to a good approximation, are [13]:

$$\omega_1^2 = \omega_{Lx}^2 \sin^2\theta + \omega_{Lz}^2 \cos^2\theta$$

$$\omega_2^2 = \omega_{Tx}^2 \sin^2\theta + \omega_{Tz}^2 \cos^2\theta \qquad 18$$

The approximation used here can be termed the quasi-cubic approximation. Lattice dispersion continues to affect both LO and TO frequencies:

$$\omega_{Li}^2 = \omega_{LOi}^2 - \Omega_{Li}(q^2) \qquad 19$$

$$\omega_{Ti}^2 = \omega_{TOi}^2 - \Omega_{Ti}(q^2)$$

Here $\omega_{LOi}$ and $\omega_{TOi}$ are the zone-centre frequencies. There is, thus, a TO band and a LO band. The latter is of interest regarding the interaction with electrons; as $\theta$ increases, the frequency of the mode shifts from $\omega_{Lx}^2$ to $\omega_{Lz}^2$ and from $\omega_{Tz}^2$ to $\omega_{Tx}^2$; along with that, the dielectric constant shifts from $\kappa_x$ to $\kappa_z$. For a given frequency, the permittivity can be written:

$$\epsilon(\omega) = \epsilon_\infty \frac{\omega^2 - \omega_1^2}{\omega^2 - \omega_2^2} = 0 \qquad 20$$

As regards the IF mode,

$$\kappa_x \sin^2\theta - \kappa_z \cos^2\theta = 0 \qquad 21$$

With $q_x^2 + q_z^2 \approx 0$, plus the cubic approximation, eq.21 is satisfied for arbitrary frequency. Thus, a hybrid can exist with LO and IF components. Where the frequency of the IF mode is the same as the frequency of the LO mode, whatever the dispersion.

### 3. **The LO, IF and TO Hybrid**

A hybrid optical mode can be formed with LO, IF and TO components with common frequency and common $q_x$ that satisfies both the mechanical and electrical boundary conditions at an interface. For example, in the case of a quantum well such as would occur in an AlN-GaN-AlN structure, in which the quantum well occupies $-L/2 \leq z \leq L/2$, the symmetrical components of the ionic displacement are:

$$u_x = q_x A e^{i(q_x x - \omega t)}[\cos q_z z + B \coth q_x z + C \coth \eta z] \qquad 22$$

$$u_x = q_z A e^{i(q_x x - \omega t)}[D \sin q_z z + F \sinh q_x z + G \sinh \eta z$$

There will also be an antisymmetrical hybrid with $\cos \to \sin$ $and$ $\coth \to \sinh$. A is an amplitude to be determined by energy normalization, and is assumed regarding this equation, that $\omega > \omega_{TO}$, so that the TO component is an interface mode. The LO and IF components are longitudinally polarized, so that $\nabla \times \boldsymbol{u} = 0$, and the TO component must obey $\nabla \cdot \boldsymbol{u} = 0$. These conditions, plus the requirements of mechanical and electrical continuity at the interface, determine the other amplitudes. For simplicity we have regarded the basal plane as isotropic and the x direction arbitrary, so that all components travel along the interface in the same direction and with the same wave vector ($q_x$) and the same frequency.

The mechanical boundary conditions are complicated by the fact that there is a difference of reduced mass across the interface with a corresponding difference of frequency. Commonly adopted for simplicity is the effective boundary condition **u**=0 at the interface. Perpendicular to the interface all waves in the quantum well are confined and stationary.

As is well-known, in the interaction with electrons, momentum and energy are conserved. This produces limits on to the range of $q_x$ which applies to the necessary integration over $q_x$. The sum over $q_z$ is affected by confinement and, in the case of hexagonal crystals, the dependence of frequency on the direction of propagation according to eq.18.

It is worth pointing out that in the extreme long-wavelength range of LO modes the effect of dispersion is negligible. In this case, the mechanical and electrical boundary conditions can be satisfied to a good approximation without the presence of a TO component. Thus:

$$u_x = q_x A e^{i(q_x x - \omega t)}[\cos q_z z - B \coth q_x z] \qquad 23$$

$$u_x = iq_z A e^{i(q_x x - \omega t)}[\sin q_z z + (q_x/q_z) B \sinh q_x z]$$

$$B = \frac{cos q_z L/2}{cosh q_x L/2}$$

$$tan q_z L/2 = -\frac{q_x}{q_z} tanh q_x L/2$$

No fields appear in the barrier. Equally, no barrier fields appear in the well. In the extreme long-wavelength limit there are no remote-phonon effects. The latter can be seen as a consequence of lattice dispersion.

Energy normalization and quantization are modelled on the simple harmonic oscillator as usual. The energy of the hybrid is the sum of mechanical and electrical. As a travelling wave along the interface, its potential and kinetic energy components are equal, so the total energy density can be written:

$$H = \frac{\omega^2 \mu}{V_0} u^2 + \epsilon E^2 \qquad 24$$

The LO component has $\epsilon = 0$, so its energy is purely mechanical, as is the energy of the TO component. The IF component is the only component with electrical as well as mechanical energy. The electrical energy is:

$$\epsilon E^2 = \frac{(\omega^2 - \omega_L^2)(\omega^2 - \omega_T^2)}{(\omega_L^2 - \omega_T^2)} \frac{\mu}{V_0} u^2 \qquad 25$$

The ratio of electrical to mechanical energy-density is:

$$\frac{(\omega^2 - \omega_L^2)(\omega^2 - \omega_T^2)}{\omega^2 (\omega_L^2 - \omega_T^2)} \qquad 26$$

Clearly, this ratio is small, and the electrical energy is usually neglected. Nevertheless, the IF mode has mechanical energy, as has been recognized, for instance, in the model of Mori and Ando [14], which was based on the dielectric continuum model. Energy normalization is therefore adequately carried out assuming the existence of mechanical energy only.

### 4. Comments

We note that the variation of LO frequency with direction of propagation makes the calculation of the electron-phonon interaction tedious, especially with regard to the use of the ladder technique needed to assess the momentum-relaxation rate. The assumption of the quasi-cubic approximation - $\omega_{LOx} \approx \omega_{LOz}, \omega_{TOx} \approx \omega_{TOz}$, - suggests that a cubic approximation would be useful for determining the strength of the electron-phonon interaction where extreme accuracy is not needed.

### 5. Conclusions

The foregoing theory of hybrid phonons in a hexagonal crystal is sufficient to calculate the interaction between polar modes and electrons for any nanostructure composed of hexagonal crystals where extreme accuracy is not required. Besides the limitations concerning long wavelengths and the neglect of lattice dispersion, which it shares with the DC model, its relative simplicity relies on the difference between LO frequencies and the difference between TO frequencies in the two directions of interest to be small enough to be negligible, what we have termed the quasi-cubic approximation. As a result, well-defined frequency bands for LO, and TO modes can be identified. For the same reason, it makes it possible to obtain a quick estimate by regarding the crystal as approximately cubic –a full cubic approximation, in short.

One of the main differences with the DC model is the elimination of any electrical influence of longitudinal IF modes in the well, also the elimination of the influence of IF modes in the barrier material. In the long-wavelength model there is no remote-phonon effect. Going beyond the long-wavelength approximation would require a comprehensive account of the lattice dynamics and the effect of mixing of optical and acoustic modes [15].

Acknowledgement

We would like to thank the US Office of Naval Research for their support via Grant Nos. N00014-18-2373 & N00014-18-1-2463 sponsored by Paul Maki.